

\magnification=1200
\hsize=15.2truecm
\vsize=21.6truecm
\baselineskip=14pt

\vglue 1truecm
\centerline {{\bf SOFT PHYSICS AND INTERMITTENCY : }}
\centerline {{\bf OPEN QUESTION(S) in KRAKOW}}
\vglue 1.5truecm
\centerline {R. Peschanski}
\centerline {{\it Service de Physique Th\'eorique, D.S.M.}}
\centerline {{\it CEA-Saclay, 91191 Gif-sur-Yvette Cedex, France}}

\vglue 1.3truecm

\centerline {ABSTRACT}

{\leftskip=1.2truecm \rightskip=1.2truecm
This contribution contains  a summary of the Krakow
meeting on Soft Physics and Fluctuations. It emphasizes both
the experimental and the theoretical
investigations of correlations/fluctuations and
intermittency in multi-particle
processes and discusses of the present status of this concept. A clarification
of the main open
questions in this field of research is now within reach, thanks to the studies
presented at
the meeting.
 \par}
\vskip 0.5truecm
\noindent {\bf Introduction}
\vskip 0.5truecm

Let me start by an introductory warning: one of our contributors to the meeting
$\lbrack$1$\rbrack$, has discussed
 the existence of an {\it unbiased}
estimator for dynamical fluctuations (we will return more seriously to
this important topics). To be clear, this summary is not made by an {\it
unbiased}
estimator! There are at least two reasons for that. First, being a theoretician
I have
no competence to evaluate the validity of experimental
results. The best thing I can do is to propose a suitable observable and
ask for advice from my experimentalist friends. Second, theory is a ground for
subjectivity. One way towards objectivity is to take care of
everybody's work. I will thus try my best! In fact, I will only use
the contributions to the workshop, and their remaining tracks in the
proceedings,
for the discussion and references.

During the meeting there were nice contributions to the so-called
"Soft Particle Physics" which do not concern fluctuations/correlations in
multiparticle production. Section {\bf 1} of this talk is devoted to these
aspects. Experimental
aspects of the search for soft photons, theoretical contributions to the
search of a non-perturbative approach to Quantum ChromoDynamics (QCD)
enabling to
describe "Soft Physics", the use of studies based on nuclei, were all
important features of our meeting, and will be shortly
reviewed.

The correlation/fluctuation question and related studies
occupied a large fraction of the meeting. The main reason for this
are the stimulating discussions that this field of research
has been leading to during the last few years, especially in relation with
the concept of "Intermittency". Indeed, many of the unclear issues
have now been clarified, but, all in all, not solved. That is the reason
why the next sections of this summary talk are devoted to
the most prominent results and open questions which appeared during
the meeting.

In section {\bf 2}, I will refer to the rapid evolution of the experimental
techniques
involved in the     study of correlations/fluctuations in multiparticle
physics.
The comparison of the tools available nowadays with the ones originally
proposed shows the
technical advances made since then. In section {\bf 3}, we will focus on three
main
collective results which were obtained thanks to these techniques:i) the
emergence of
Bose-Einstein interference effects in the study of fluctuations, ii) the
comparison with
perturbative QCD calculations obtained by different groups for the first time,
and iii) the
intermittent behaviour in connection with phase transitions .

Section {\bf 4}
is devoted to a series of questions left
for future work which  I noticed
open during the meeting.
 May be, some of them
will find their answer by mail return,...or before the next multi-particle
meeting.

\vskip 0.5truecm
\noindent {\bf 1. Problems in Soft Particle Physics}
\vskip 0.5truecm

"Soft Physics", at least when concerning elementary particles,
appears really as the "Hard Problem". Indeed, the processes involving
hadrons with small transverse momentum between each other and/or
with respect to the incident ones remain a theoretical mystery. While such
processes have been experimentally studied since a long time, it seems that
their theoretical understanding remains at a standing point, except perhaps for
the "static" QCD calculations of the low-lying states of the
hadronic spectrum. We remain
thus far from a deep understanding of "soft" hadronic reactions at high energy
 which implies a knowledge of the non-perturbative
behaviour of Quantum Field Theory. More precisely the confinement problem
in QCD, which requires a non-perturbative vacuum shift from the elementary
quark and gluon fields to hadrons, is for the moment beyond reach. So, what to
do?

In this general context, one could think that multi-particle processes is
an
even tougher problem
in Soft Physics.
They involve a lot of particles, and thus a lot
of variables for the description
of the final states, which seems to forbid any reasonable treatment of
the scattering amplitudes in terms, say, of Feynman diagrams in the
perturbative expansion. However, less superficially, one may realize that
there are two hidden advantages which could
help the understanding on a basic level.
First, particles are to be considered as elementary quanta
of a relativistic quantum field, and thus operating with many particles could
be a better revelator of the quantum field structure than
low energy and/or low
multiplicity events. Second, the number of particles involved in modern
experiments at high energy is quite high (from dozens at LEP to hundreds
at tevatron and future accelerators). This casts a bridge between
particle physics and systems with many degrees of freedom. For instance,
some powerful
Statistical-Mechanics tools
may become relevant, and their connection with Field Theory may be of
great help, as it has already
 been the case for lattice
calculations. To my mind, the goal and spirit of the
workshop is to make progress towards these directions.

On the theoretical point of view, questions related to the field theoretical
vacuum
and more specifically the QCD vacuum at large distance have been discussed
during the meeting.
In analogy with QED, some hypotheses on the behaviour of quarks in
such a vacuum have been modeled, see$\lbrack$2$\rbrack$. It is
an ambitious approach, though apparently very difficult.
Another approach is to consider an effective field theory of
hadrons, for instance in the framework of a $\sigma$-model of pions and sigma
resonances. The interesting suggestion discussed in $\lbrack$3$\rbrack$ is that
the large number of pions to be produced in future accelerators would justify
a {\it quasi-classical} field-theoretical approach. This leads to
dramatic predictions, like the formation of domains of disordered
chiral condensates(DCC) following a very high energy collision. In more
practical
terms, one could observe bunches of particles with definite charge, e.g.
neutrals.
This picture may remain a theoretician dream, but it is worth investigating
its consequences. It has the merit to
show that imagination and theoretical rigor can live together in this domain
of research. Several studies are now in action about DCC, and one
is waiting for their results with curiosity.

On the phenomenological ground of soft physics (excluding
correlations /fluctuations) soft photon physics and the
interaction of particles with nuclei, (or of nuclei with nuclei)
were the topics represented in the meeting. Soft photons
can be considered as the Loch Ness Monster (or the legendary Krakow dragon!)
of particle physics, since they are claimed to
appear from time to time. More precisely,
the question remains to know whether a significant excess of {/it direct}
photons of low
energy is produced with respect to the known background, mainly due to
bremsstrahlung. The results shown to us by the WA83 collaboration at CERN, see
$\lbrack$4$\rbrack$, give evidence for a strong excess, but amazingly similar
in shape
and properties with the bremsstrahlung. It is clear that this interesting
experimental
finding has to be confronted with other ones wich give
negative results. Let this research
contribute to solve a long-standing controversy on this important question.
A theoretical discussion can be found in$\lbrack$5$\rbrack$.

Nuclei, beside their own interest as quantum systems of hadrons provide
valuable tools for
particle production: they may represent the finest existing microdetectors
of sizes below $10$ fermis. In particular
they may give some information on the first stages of hadron production, at
least if one
is able to distinguish nuclear from particle effects in a suitable way. In
$\lbrack$6$\rbrack$  a series of results and models are discussed, which can
pave the way towards
the use of nuclei as detectors. Nuclei properties by themselves are evoked in
$\lbrack$7$\rbrack$  together with the interesting experimental detection
of a possible critical phenomenon during
the multifragmentation of heavy nuclei.
 \vskip 0.5truecm
\noindent {\bf 2. Tools for correlations/fluctuations: Past/Present}
\vskip 0.5truecm

In order to figure out the decisive progress made in the
detection of dynamical fluctuations in multi-particle data, it is
useful to compare the tools used and discussed at the present workshop
with the original method based on
the factorial multiplicity moments, their binsize dependence and the
$\alpha$-model of
intermittency.

Factorial moments have been designed in order to remove from the
measurements of multiplicity fluctuations the statistical
fluctuations associated to a Poisson (or at
fixed multiplicity, Bernouilli) noise. This simple assumption had
the merit to exhibit unknown features of dynamical particle
correlations, which are related to these moments by construction.
Let us recall the
conventional definition of factorial moments of rank $q:$
$$
{\cal F}_{q}(\delta) \equiv {{<n(n-1)...(n-q+1)>} \over {<n^q>}}
\simeq \delta^{-f_q},   \eqno (1)
$$
where $n$ is the multiplicity of particles observed in a
phase-space interval $\delta$. The third term of the equality
represents by definition the intermittent behaviour characterized
by a set of indices $f_q$. Note that the factorial moments can
also be expressed in terms of integrals of the $q-$correlation
functions integrated in $\delta.$

When experimentalists became interested
in the game, it was recognized that the method could and should be
improved. In particular, the factorial moments suffered from
one important  defect: they were rather unstable at small binsize and high
rank. At our meeting, these problems have been discussed, see
$\lbrack$1,8$\rbrack$, and one method has been found
providing a good improvement: the correlation
integrals' method. The correlation integrals, which appear under different
forms,
have in common the following recipe: contrary to factorial moments, where
the q-uple groups of particles are counted in phase-space boxes
defined {\it a priori}, the
correlation integrals count all q-uple groups
within a given distance $\delta$. Removing the arbitrariness of the phase-space
division
gives a much stronger stability to the results. My conclusion, which
is perhaps the main positive conclusion about the meeting, is that, after a few
years of hesitation, one
disposes of good tools to evaluate dynamical fluctuations/correlations
in multi-particle physics.

As noticed since their proposal, factorial moments do not give a dynamical
information on multi-particle fluctuations unless one varies the bin-size.
However, different {\it technical} difficulties appeared when this binning
has to be done in the fully-dimensional phase-space: dependence of the
moments on a non-homogeneous average multiplicity, instabilities of
different kinds, etc... As a result of the improvements on the measure
of fluctuations, some important empirical results have been found on
the $2-$ or $3-$ dimensional cases. Let us quote in particular the
universality of moments' behaviour
 in this last case, which has been confirmed
at the meeting$\lbrack$11$\rbrack$.

The $\alpha-$model of intermittent fluctuations has been useful to
model out genuine intermittency properties in multi-particle
physics. As a mathematical tool, it allows to confront experimental data
or phenomenological models to typical sets of fluctuations/correlations without
scale (self-similar). However, it is a very crude type of modelization
when compared to the sophistication of data. At least in two cases, it has been
modified or improved. First, intermittent fragmentation models
have been worked out$\lbrack$12$\rbrack$, which possess both {\it local}
intermittent structures and {\it global} features of the multiplicity
distributions as seen in high-energy reactions. Second, Monte-Carlo simulations
of great
technicity have incorporated intermittent correlations$\lbrack$13$\rbrack$. Let
us however remark that the level of accuracy obtained in Monte-Carlo
simulations for $e^+-e^- -$reactions$\lbrack$9$\rbrack$ do not exist in other
cases. Indeed, the problem
here is due to the mismatch between intermittency and the
Bose-Einstein correlations for hadron-induced reactions, which is
discussed later on.

With the noticed improvement of tools for exploring the correlations
/fluctuations,
It is now time to go to the basic question we want to adress,
namely the physical origin of the observed dynamical fluctuations, which
 are compatible with the intermittency behaviour (1).
\vskip 0.5truecm
\noindent {\bf 3. What is the origin of intermittency?}
\vskip 0.5truecm

Good  or bad, the fact is that no
convincing
explanation of the intermittent phenomenological structure of
dynamical fluctuations in multi-particle reactions exists. Partly, it is due
to the lack of precision data for some time. It is only recently
that the emergence of the Bose-Einstein correlations
in intermittency studies have been
confirmed. Partly, it is due to the lack of theoretical
understanding of long-distance effects in field theory, which is a
subject treated during the workshop. There was also discussed an
interesting connection between intermittent fluctuations and phase transitions.
Let us
review these subjects in turn.

i) {\bf Bose-Einstein correlations}

The main finding concerning intermittency in the past year is that
it is dominated by the same charge correlations between same-sign
particles
in very small bins of the whole $3-$dimensional phase space
at least for hadron or nuclei-induced reactions.
This year,
the evidence for the contribution of same-sign particle correlations
have been confirmed and strengthened in various cases, namely in
hadron-hadron reactions examined by the NA22 experiments at $22
GEV/c$$\lbrack$10$\rbrack$
and UA1 at $640 GEV/c$$\lbrack$14$\rbrack$. However, the situation is different
in
$e^+-e^-$ annihilation into hadrons as seen, e.g. by DELPHI at
LEP$\lbrack$9$\rbrack$. Indeed, the correlation integral method clearly shows
that for
hadron-induced reactions,
same-sign particles contribute mainly at very small $q^2$, the
Lorentz-invariant momentum distance between near-by particles.
Note that a check should be
done to know whether this variable gives the same results as the $3-$
dimensional phase space in terms of rapidity, azimuth and
transfer momentum. For lepton-induced
reactions, the opposite is true, namely correlations seem to be due to
opposite-charge particles$\lbrack$14$\rbrack$. The situation in the "mixed"
case,
that is lepton-hadron interactions is also "mixed"! Indeed,
same-sign
particles give the main contribution $\lbrack$15$\rbrack$ at CERN
energies, but are dominated by other effects in Monte-Carlo simulations
at HERA energies, probably due to gluon cascading, see further on.

Does it mean that the conventional Bose-Einstein interference effect between
identical particles can explain the intermittency phenomenon? Or what
is the influence of Bose-Einstein correlations on the intermittency problem
(as it is asked in ref.$\lbrack$14$\rbrack$)?  The investigations
described
during the meeting bring some interesting precisions on the problem. First,
the form of Bose-Einstein
correlations have been studied in detail, starting from the conventional
gaussian or di-gaussian fits$\lbrack$14,16$\rbrack$. Quite unexpectedly,
smaller
is the interval in $q^2$ available, more peaked is the form of the correlation
curve. One goes from
gaussian to exponential
$\lbrack$14,16$\rbrack$, or to an edgeworth expansion (using Hermite
polynomials)
$\lbrack$17$\rbrack$, or finally a power-law form without intrinsic scale
for the source radius$\lbrack$10,14$\rbrack$. This last case, though
required only by the
very forward region of correlation, is the result
obtained via the factorial or correlation integral methods. If confirmed by a
detailed
analysis of these very near-by correlations, this
would mean a scale-invariant behaviour of the Bose-Einstein mechanism
itself.

On the theoretical point of view, while waiting for a development of the
experimental
situation on this subject, it is stimulating to examine the motivations for
having such a scale-invariant structure of the
Bose-Einstein correlations. This has been analyzed in
$\lbrack$18$\rbrack$, with the following
conclusions: in any case there must be an (effective) singularity in the
space-time structure of the source of correlations (effective: it acts as a
singularity
in some region near-by, with a cut-off to avoid a true, unphysical, singularity
in the cross-sections). Now, two cases, at least, are possible: the source is
event-by-event smooth (not fractal),
 but it fluctuates with a singular distribution from event to event; Or,
a source is not characterized by a smooth curve , but is itself a fractal
object in space-time,
and the scale-invariant behaviour is then a consequence of the scale-invariant
structure of the dynamics (as in classical intermittency). One problem
on previous studies raised by this approach is the lack of solid derivation of
the Bose-Einstein
correlations when there exist correlated sources, since the
Hanbury-Twiss phenomenon is for uncorrelated sources. A key phenomenological
conclusion of this work is the necessity to look for higher-order Bose-Einstein
correlations (with more than 2 particles involved) as a way of distinguishing
the various mechanisms.

ii){\bf Perturbative QCD predictions}

We have seen that in $e^+\!-e^-$-annihilation into hadrons, the Bose-Einstein
effect is likely
not to be
the dominant mechanism for scale-invariant fluctuations, even in full
dimensionality. In this case, one can be confident that, at least in a first
stage of
the reaction, perturbative QCD calculations can give a hint on
to the problem. At a deeper level,
it is a basic problem to see whether the intermittency effect has some
connection with
the fundamental theory at all! Interesting developments have been reported at
the conference
$\lbrack$12,19$\rbrack$, showing from three different calculations
that perturbative QCD, within some approximation framework, is
intermittent in the strict sense for emitted gluons (in the axial gauge).
However, hadronization effects may be important,
since the predicted behaviour for gluons is similar in form but different in
strength
from that observed for hadrons$\lbrack$19$\rbrack$.

Some conceptual and computational problems had to be solved before arriving at
the
results, which may explain why this well-defined problem
in perturbative field theory took a long time and many efforts before
yelding a solution.
Conceptually, it was difficult to understand how a perturbative theory at short
distance can
say something for a typically long-distance problem. However, handling all
orders of the perturbative expansion
with the leading logarithm approximation, and the choice of a well-adapted
gauge,
allows one to make predictions for the multiplicity of gluons (quasi-real in
such a gauge-fixing) in a small phase-space interval. On the computational
point-of-vew, it was rather striking that,
within the same approximation scheme, analytical results on the behaviour of
factorial moments could be obtained.

The overall result, besides some differences probably due to different
assumptions
on the initial conditions and on the observables, is that fractal dimensions
can be predicted from these calculations. Let us first, for the sake
of simplicity, fix
the coupling constant of QCD. Then the behaviour of fluctuations
is exactly fractal, that
is effective singularities of the $q-$correlation functions exist and are
concentrated into fractal regions of
phase-space, which are random but with fixed non-integer dimension
$d \equiv {f_q \over q-1}$
, see equation (1). When the
running of $\alpha_{S}$ is restored a more complex behaviour appears with
multifractality and saturation at small bins. Multifractality means that now
the
$q-$correlation functions have $q-$dependent dimensions. Saturation corresponds
to the
breakdown of the intermittent behaviour since the amount of gluon radiation
becomes
so large that the fractality disappears and the full phase-space is filled by
the
gluons. Note that this breakdown appears already at the perturbative level,
quite unexpectedly. However, for various reasons,
the small-bin region becomes ambiguous and in fact forbidden to
perturbative calculations. We are entering the no-man's land of field theory
(for
the moment, I hope,) the non-perturbative regime. A proposal is
made for damping hadronization effects by looking for ratios of factorial
moments with varying resolution
$\lbrack$12$\rbrack$.
Concluding with that subject, there is now a theoretically motivated route to
intermittency in
QCD which deserves more study. The problem of non-perturbative methods remain
completely
open. Note, however, the interesting attempt of $\lbrack$20$\rbrack$,
introducing a new mechanism within the popular string fragmentation picture of
hadronization.

iii) {\bf Phase Transitions}

As is well established by the recent developments in field theory,
when you cannot solve a non-perturbative problem,
you discretize your problem  on a lattice and you study phase-transitions. This
is precisely what
people also did for the intermittency problem, connecting it to the study of
fluctuations
(both dynamical and geometrical) at a phase transition.
There has been  a significant activity in this domain using typical Statistical
Mechanics methods. On a lattice, you may easily create the conditions of a
critical behaviour leading to intermittent fluctuations, without the technical
constraint of
a weak coupling. For instance the effective Ginzburg-Landau theory of
phase transitions have been advocated$\lbrack$21$\rbrack$. The main problem
of this kind of studies is that one is not sure to meet the
requirements for a "pure" phase transition at equilibrium for multi-particle
production. Even in the case of heavy-ion
reactions a thermal transition from the quark-gluon plasma is not
an evidence. On the contrary, the observed fluctuations in this
case seem to be quite different from those expected from the
formation and decay of a plasma.
I am nevertheless confident that the richness and the variety of physical
situations which one
meets in Statistical Mechanics systems will give powerful tools to particle
physicist in
the near future, for instance considering non-equilibrium systems.
This is a guess.
\vskip 0.8truecm
\noindent {\bf 4. Concluding by questions}
\vskip 0.8truecm

{}From the preceeding discussion, it is clear that the phenomenons, registered
under the name of intermittency in
multi-particle reactions, have not yet found a physical interpretation. While
in
$e^+\!-e^-$ reactions into jets, the hadronization contribution is not
understood
and remains ambiguous in strength, see e.g.$\lbrack$22$\rbrack$, the interplay
of Bose-Einstein correlations with a possible scale-invariant structure of
interactions
remain a mystery in the other cases. It is thus too soon to draw any definite
conclusion. Better is to propose a series of questions for further
investigations.
The interest of the situation is that, thanks to the sizeable improvements of
the experimental and theoretical tools, some of the  answers
to the open questions could be
within reach in the near future.

{\bf 1. Wavelets and other methods.}
Can we develop new experimental tools, such as
wavelets$\lbrack$1$\rbrack$, in order
to even more improve the data on correlations /fluctuations? The key question
seems to be the possibility of mixing these probabilistic methods with the
"factorial trick" in order to avoid analyzing only the statistical "noise" by
wavelets.

{\bf 2. The "Wall and Tower" problem.}
As is clear, e.g. in the analysis of ref.$\lbrack$14$\rbrack$, the factorial
analysis in different dimension can reveal different types of dynamical
excitations.
A "wall" of particles represented in a Lego-plot could be seen by perpendicular
projection
in phase-space as a strong fluctuation, while a "tower" is likely to dominate
only the
full dimensional phase-space. If then, Bose-Einstein correlations dominate in
most cases
the $3$-dimensional studies, what about lower dimensionalities? In particular
how to interpret the Ochs-Wosiek scaling (moments over moments) in $1$- and
$2$-dimensional
NA22 studies$\lbrack$10$\rbrack$?

{\bf 3. The "multiplicity and $P_T$" problem.}
In hadron-induced reactions, it has been noticed$\lbrack$14$\rbrack$ that, in
contradiction with lepton-induced reactions,
 simulations are far from reproducing the low multiplicity and $P_T$ behaviour
of
fluctuations. As such this remains an unexplained feature which,
if a solution is found, can open some
doors for the "soft" part of the intermittency problem. One should notice in
this
respect the new Monte-Carlo simulations, with
intermittent fluctuations, applied to soft physics$\lbrack$19$\rbrack$.

{\bf 4. The "Universality" problem.}
It has been noticed that $3-$dimensional factorial moments follow
a quite striking general behaviour in various reactions, such as lepton-hadron,
hadron-hadron and even heavy-ion ones$\lbrack$11$\rbrack$. More precisely,
the second factorial cumulant $K_2$ ($\equiv F_2  -1$) has roughly the same
scale-invariant
behaviour in all cases, up to a constant which can be attributed to the
different
long-range correlations. Is this feature due to the common Bose-Einstein origin
of the expected fluctuations? Could we expect such a large extension of the
scale-invariant range (from 1 to $10^4$ subdivisions) in all cases?

{\bf 5. "Angular intermittency".}
One property of intermittent fluctuations in particle physics suggested by
perturbative QCD calculations is that an effective singularity
may show up in the fluctuations/correlations for angular
variables$\lbrack$12$\rbrack$. It would be interesting to look
for such a behaviour, either in lepton-induced reactions by depressing as much
as possible the effect of hadronisation$\lbrack$12$\rbrack$, or by extension
to all other cases, just for curiosity's sake.

{\bf 6. Intermittency at HERA?}
 An interesting remark made during the conference$\lbrack$15$\rbrack$, is that
one expects in lepton-hadron reactions (deep-inelastic scattering) a
competition
between the Bose-Einstein type of correlations and the perturbative QCD-induced
 mechanism. More precisely Monte-Carlo simulations indicate a levelling off of
the perturbative component with the energy. This gives hope that the
properties of the underlying perturbative theory can be caught
from the
fluctuations/correlations observed in the final hadron radiation.
Such a study may be of some help in discussing such matters as
the improved perturbative expansions(cf. the Lipatov regime), coherence of
gluons
in the space-like region or saturation effects of partons beyond the
perturbative
regime? This is a challenge for next future.

{\bf 7. Intermittency and the space-time structure of Strong Interactions.}
On the theoretical side of the problem, a much
better understanding of the space-time development of the processes is
required$\lbrack$18$\rbrack$. The extension of the interactions
to large distances, which is probably necessary to generate scale-invariant
fluctuations, is the origin of most (if not all) open questions: what is the
interplay between
parton fragmentation, resonance production and decay, quantum interference
effects,
dynamical quark-gluon phase transitions, in the overall phenomenon?
Is there a simple answer or do we face the physics of a complex system? Those
are
some of the basic questions one has to face in that game.

{\bf 8. The "Micro-Universe story".}
Let us end by the a speculative touch; Why not dream sometimes?
The problematics of intermittency leads to an analogy between the history of
the macro-world
and that of the  micro-world. In the macro-world, e.g. the Universe, after the
Big-Bang,
there was a succession of self-organizing structurations, compensated
(as the entropy increases) by an
increasing disorder whose signature is the famous Background Radiation. In the
Micro-Universe represented by a production of particles from the vacuum, it is
tempting to find the origin of dynamical fluctuations in the structuration
of partons during the scattering process. They tend to form colorless clusters,
but
they do not necessarily find their partners within the small range of their
individual interaction.
Fluctuationg structures are formed, rearrangements occur, till they succeed
to form new objects, the hadrons, which are required by the new vacuum
structure.
These hadrons seem to be rather complicated and structured objects at low
energy,
which may explain the long duration of the structuration compared with the
length of the fundamental interaction. Is this picture right
or wrong?
\vskip 0.8truecm
\noindent {\bf 5. Many Thanks To The Organizers!}
\vskip 0.8truecm
It is a pleasure to warmly thank all the organizers of our meeting.
The atmosphere of discussions, exchanges and friendship show
the evidence that
the goals of the conference have been realized much beyond the normal level.
Thanks to all of them.

\vskip 0.8truecm
\noindent {\bf Contributions' References in the Proceedings}
\vskip 0.8truecm

\item{$\lbrack$1$\rbrack$} {\bf P. Lipa.}

\item{$\lbrack$2$\rbrack$} {\bf O. Nachtmann} , {\bf W. Czyz} .

\item{$\lbrack$3$\rbrack$} {\bf J.-P. Blaizot} , and references therein.

\item{$\lbrack$4$\rbrack$} {\bf M. Spyropoulou-Stassinaki} , {\bf T. J.
Brodbeck} .

\item{$\lbrack$5$\rbrack$} {\bf W. Florkowski} , {\bf J. Pisut} .

\item{$\lbrack$6$\rbrack$} {\bf P. Hoyer} , {\bf J. Czyzewski} , {\bf E.
Stenlund} .

\item{$\lbrack$7$\rbrack$} {\bf M. Ploszajczak} , {\bf R.P...}

\item{$\lbrack$8$\rbrack$} {\bf H. Egger} s and references therein.

\item{$\lbrack$9$\rbrack$} {\bf F. Mandl} .

\item{$\lbrack$10$\rbrack$} {\bf W. Kittel} .

\item{$\lbrack$11$\rbrack$} {\bf K. Fialkowski} .

\item{$\lbrack$12$\rbrack$} {\bf J.-L.  Meunier.}

\item{$\lbrack$13$\rbrack$} {\bf E. de Wolf} , {\bf R. Hwa} , {\bf J. Rames} .

\item{$\lbrack$14$\rbrack$} {\bf B. Buschbeck}  and references therein.

\item{$\lbrack$15$\rbrack$} {\bf I. Derado} .

\item{$\lbrack$16$\rbrack$} {\bf P. Seyboth} .

\item{$\lbrack$17$\rbrack$} {\bf T. Csorgo} .

\item{$\lbrack$18$\rbrack$} {\bf A. Bialas} , see also {\bf B. Ziaja} .

\item{$\lbrack$19$\rbrack$} {\bf J. Wosiek} , {\bf W. Ochs} .

\item{$\lbrack$20$\rbrack$} {\bf G. Gustafson} .

\item{$\lbrack$21$\rbrack$} {\bf R. Hwa} .

\item{$\lbrack$22$\rbrack$} {\bf J. Rames} .

\vfill
\eject
\bye